\documentclass[a4paper]{article}

\usepackage{newtxtext} 
\usepackage{newtxmath}
\usepackage[T1]{fontenc}
\usepackage{multicol}
\usepackage{titlesec}
\usepackage{enumerate}
\usepackage{enumitem}
\usepackage{booktabs}
\usepackage{array}
\usepackage{hyperref}
\usepackage{fancyhdr}
\usepackage{tabularx}
\usepackage{lipsum}
\usepackage{graphicx}
\usepackage{cleveref}
\usepackage{etoolbox}

\hypersetup{
    colorlinks=true, % Enables colored links
    linkcolor=red,   % Color for internal links (e.g., \ref, \cite)
    urlcolor=blue,   % Color for URLs (e.g., \href)
    citecolor=green,  % Color for citations
    pdfborderstyle={/S/U/W 1} % Underline style: solid, underline, width 1pt
}
\setlist[itemize]{leftmargin=*,topsep=4pt,itemsep=0pt,parsep=0pt,partopsep=0pt}

\titleformat{\subsection}      % run-in heading (inline)
  {\itshape}                      % italic for number + title
  {\thesubsection.}               % subsection number with dot
  {0.5em}                         % space between number and title
  {}                              % code before the title
\titlespacing*{\subsection}{0pt}{1ex}{1em} % small spacing before/after
\titleformat{\subsubsection}[runin]
  {\itshape}                      % italic for number + title
  {\thesubsubsection.}            % subsubsection number with dot
  {0.5em}                         % spacing
  {}
\titlespacing*{\subsubsection}{0pt}{1.5ex plus .1ex minus .2ex}{1.5ex plus .2ex}

\usepackage[
    top=1in,
    bottom=1in,
    left=0.59in,
    right=0.59in,
    headheight=61pt, % Sufficient height for logo/text
    headsep=0.2in, 
    footskip=0.49in
]{geometry}

\usepackage{caption}

\makeatletter
\def\ciredmaketitle{
    \vspace*{-1 cm} 
    
    \begin{center}
        {\fontsize{18pt}{22pt}\selectfont \bfseries \@title \par} 
        \vspace{0.3cm} % Space between title and author
        
        {\lineskip 0.5em 
        {\fontsize{12pt}{14pt}\selectfont \bfseries \itshape \@author \par}
        }
        
        \vspace{-.9 cm} % Space between authors and affiliations
        
        \@date \par
    \end{center}
    \par
    \vspace*{0.5cm} % Space after the title block before the main content
}
\makeatother
\begin{document}

% 4a. Define Title, Author, and Affiliations
%\title{Designing Active Operation in Low-Voltage Distribution Grids: Requirements, Interfaces and Roadmap}
\title{DESIGNING ACTIVE OPERATION IN LOW-VOLTAGE DISTRIBUTION GRIDS: REQUIREMENTS, INTERFACES AND ROADMAP}
\author{Eric Tönges$^1$$^*$, Andrea Schoen$^{1,2}$, Frank Marten$^{2}$, Marco Pau$^{1,2}$, Denis Mende$^{1,2}$}
\date{} 

% Define the Affiliations content using a new command

% 4b. Use the custom command to generate the title and author block
\ciredmaketitle 
 \begin{center}

    $^1$University of Kassel, Kassel, Germany \\
    $^2$Fraunhofer Institute for Energy Economics and Energy System Technology IEE, Kassel, Germany \\
    % $^3$Third Affiliation name, City, Country \\
    % $^4$Current affiliation name, City, Country \\
    $^*$eric.toenges@uni-kassel.de%

\end{center}
\thispagestyle{fancy}
 \begin{center}
  \end{center}
\begin{multicols}{2}

\section*{Abstract}
This paper outlines a pathway towards active operation of low-voltage distribution grids. In these grids, the growing deployment of distributed generation, controllable demand and storage, together with the roll-out of intelligent metering systems, creates new requirements and opportunities for distribution system operators. On the basis of the German and European regulation, and in particular of recent directives enabling grid-oriented interventions and market-based procurement of flexibility, the paper identifies three key pillars for active low-voltage operation: (a) measurement placement and observability, (b) secure and interoperable information and communication architectures and interfaces, and (c) integration of market-based and grid-oriented optimisation for controlling connected assets. A structured system overview is developed that specifies main actors and data flows, highlighting central research topics across these pillars. 
Building on this, a four‑phase roadmap is presented, spanning requirements and use‑case definition, method development and simulation, laboratory and field validation, and roll‑out with system‑level feedback, thus providing guidance for distribution system operators and researchers.
%Building on this, a four-phase roadmap is presented that spans from requirements and use case definition through method development, simulation to laboratory, field validation and roll-out with system-level feedback, providing guidance for distribution system operators and researchers.
% The abstract should be suitable for direct inclusion in abstracting services as a self-contained article. The length of the abstract \textbf{should not exceed 200 words}. Do not include figure numbers, table numbers, references or displayed mathematical expressions
%\end{multicols}

%\begin{multicols}{2}
\vspace{-0.2em}
\section{Introduction} \label{sec:introduction}
In low-voltage (LV) distribution grids, the increasing deployment of distributed generation, controllable demand and storage, together with the planned rollout of intelligent meters and Smart Meter Gateways (SMGWs), enables DSOs to mobilise flexibility. As these deployments scale, active control of connected assets becomes standard practice and a central operational objective.
In Germany, the Energy Industry Act (EnWG) frames LV operation via its Sections~\S14a (grid-oriented congestion management) and \S14c (market-based flexibility) \cite{EnWG,BNetzA.14a}. Delivering active operation requires coordinated interaction across grid and market actors, supported by interoperable, secure and compliant digital processes and data flows, and a concrete roadmap aligning regulation, measurement and observability, communications and data management, and market-based optimisation with grid-oriented safeguards \cite{Heid2025_Future14aEnWG}.
 
To support DSOs dealing with these challenges, three main pillars for the active operation of LV distribution grids are identified in this paper:
\begin{itemize}
    \item[  a)] placement and selection of measurements to achieve system awareness by means of state estimation,
    \item[  b)] secure, resilient and efficient information and communication architectures and interfaces,
    \item[  c)] integration of market-based and grid-oriented optimisation to control load, generation and storage systems.
\end{itemize}
A systematic analysis identifies requirements and interfaces for integrated operation and reveals existing challenges and research gaps across these three pillars. 
To this end, the regulatory and policy landscape across national and European levels is analysed, covering regulation of controllability and flexibility procurement, interoperability, data protection, and cybersecurity. 
Based on this analysis, a target system overview with distinct roles, interfaces and research fields is derived, and concrete steps for progression from theoretical development to validation in real grids are defined, paving the way towards active operation in LV distribution grids.
%A literature review then synthesizes research on observability and measurement selection, state estimation, and price-based mechanisms including variable grid charges and dynamic tariffs. 
%Monitoring reports and policy documents are examined to align targets and timelines.
%This systematic analysis helps identify requirements and interfaces for integrated operation, and detect existing challenges and research gaps across the three pillars. 
%Afterwards, a target system overview with different roles, interfaces and research fields is derived, to then define concrete steps for progression from theoretical development to validation in real grids, paving the way toward active system operation in low-voltage distribution grids.

In summary, the main contributions of this work are:
\begin{itemize}
    \item a structured overview of the regulatory framework for an active, market-based operation of LV grids in Germany,
    \item a system-level analysis identifying the main components, actors and research questions,
    \item a phased roadmap that structures the research activities and practical steps required to achieve active LV grid operation.
\end{itemize}

The remainder of this paper is organised as follows.
Sec.~\ref{sec:regulation} provides an overview of relevant regulatory frameworks and policy papers, building the regulatory and political foundation.
Sec.~\ref{sec:system} derives a structured system overview and outlines key roles, their interconnections and central research questions.
In Sec.~\ref{sec:roadmap}, these research questions are translated into a practical outline of how a market-based, active LV grid operation can be realised.
Sec.~\ref{sec:conclusion} concludes our findings and outlines future research and implementation challenges. % potential next steps for DSOs and researchers.

\vspace{-0.2em}
\section{Regulatory framework} \label{sec:regulation}
Active operation in LV grids is shaped by regulation, standards and policy. 
This section summarises key European and German instruments, structured by the three pillars (Sec.~\ref{sec:introduction}), and distinguishes legal obligations, technical and interoperability frameworks, and strategic energy policy objectives that guide DSO flexibility, data flows and secure digital processes.

In Germany, the Energy Industry Act (EnWG) is the central legal framework for electricity networks and markets \cite{EnWG}.
Section~\S14a establishes the conditions under which DSOs may curatively curtail controllable loads in LV grids to solve local congestion. 
It specifies the requirements for grid operator interventions \cite{BNetzA.14a} and for monetary compensation of consumers \cite{BNetzA.14a_BK8}.
Based on these specifications, DSOs must equip their LV grids with intelligent metering systems (IMS) and establish state estimation and control algorithms. 
% For active LV grid operation, 
This mandated technical setup should be leveraged with as little additional hardware and installation effort as possible.
While Section~\S14a EnWG currently refers exclusively to controllable loads, extensions to generation units have already been proposed by the German grid technology and operation committee (VDE~FNN) \cite{VDE_FNN_Impuls_2024}.
They also defined Technical Connection Rules for the integration of flexible loads and decentralized energy resources into LV grids \cite{VDE_LV_Rules}.
The legal basis for DSOs to procure flexibility services through market-based, transparent and non-discriminatory mechanisms is given in Section~\S14c EnWG, although detailed specifications are still pending. 
Section~\S14c defines the principle that congestion management can, in part, be organised through explicit flexibility products provided by prosumers, aggregators or other third parties, forming the foundation for market-based grid optimisation that is integrated with grid-oriented interventions under Section~\S14a EnWG.

Several market-based instruments are already in place. Variable grid fees are defined in \cite{BNetzA.14a_BK8} and allow LV grid operators to adapt charges to expected grid situations, but they are set in advance and do not enable reactions to real-time or short-term congestion. 
Section~\S41a EnWG \cite{EnWG} obliges suppliers to offer dynamic retail tariffs that reflect time-varying wholesale prices. 
Such tariffs can mobilise demand response and storage by exposing customers to price volatility, but, as they are typically based on zonal wholesale prices, they do not systematically reflect local LV constraints. 
For LV generation units such as PV plants, market participation to sell generated energy is defined in the Renewable Energies Act (EEG) \cite{EEG}, where remuneration is linked to real-time market prices and likewise provides no direct means for grid operators to address congestion in specific grids or individual feeders.

Redispatch~2.0 establishes a coordinated, largely market-based congestion management regime, primarily targeting generation units with installed capacities above 100~kW \cite{BNetzA_Redispatch_2020}. 
Its direct relevance for LV operation is currently limited, but the underlying principles of integrated, data-driven congestion management are increasingly discussed as a blueprint for future "Redispatch~3.0" approaches that could include LV flexibility \cite{RD30,BMWK_Dataflex}, where potential conflicts between market-based flexibility use in LV grids and flexibility provision for higher voltage levels must be avoided. 
At the European level, Directive~(EU)~2019/944 on common rules for the internal market for electricity \cite{ElectricityDirective2019} provides the overarching market and governance framework.

Active LV grid operation critically depends on digital infrastructure for measurement, communication and control. 
For grid- and market-oriented optimisation, the smart meter gateway (SMGW) is the key technical endpoint through which control commands (for example for Section~\S14a interventions) and tariff or tariff-related signals (for example dynamic prices or variable grid fees) are communicated to customer installations. 
In Germany, this infrastructure is primarily governed by the Metering Point Operation Act (MsbG) \cite{MsbG}.
Protection profiles and technical guidelines of the Federal Office for Information Security (BSI) \cite{BSI_Schutzprof,BSI_SMGW} specify mandatory requirements for cryptographic protection, authentication, role- and rights-based access control and standardised interfaces.

On the European level, the revised Network and Information Security (NIS2) Directive \cite{NIS2_2022}, nationally implemented through \cite{NIS2UmsuCG_2024}, introduces binding cybersecurity requirements for energy-sector entities and influences data granularity, access rights, permissible control paths and acceptable risk levels. 
For applications using artificial intelligence, the EU AI Act \cite{EU_AI_Act_2024} provides the regulatory framework for deploying AI-based methods in planning and operation.
Beyond regulatory frameworks, the monitoring report on the energy transition \cite{EWI_BET_2025} and the "ten key measures" for accelerating the energy transition \cite{BMWK_Schluesselmassnahmen_2025} identify accelerated grid expansion, enhanced system efficiency and stronger use of flexibility as central action fields. 
They underline the political relevance of active LV distribution system operation and frame the strategic context in which the regulatory and technical instruments discussed above are interpreted.

Table~1 summarises the discussed instruments across the three pillars in three categories: binding regulation, technical and interoperability standards and policy targets or strategies. 
The matrix highlights that each pillar is shaped by legal provisions, technical regulation and policy objectives, and that no single instrument solely governs active LV grid operation.\\

Table 1 Overview of regulatory and technical frameworks\\
\\
\begin{tabular}{
  >{\centering\arraybackslash}p{2.5cm}
  >{\centering\arraybackslash}p{1.5cm}
  >{\centering\arraybackslash}p{1.5cm}
  >{\centering\arraybackslash}p{1.5cm}
}
    \hline
       & Binding regulation & Standards / research & Policy / strategies\\\hline
       a) Measurement \& observability& \cite{EnWG,ElectricityDirective2019,MsbG,EU_AI_Act_2024}  
       & \cite{VDE_LV_Rules,BSI_Schutzprof,BSI_SMGW}  
       & \cite{EWI_BET_2025} \\
       b) Communication \& interfaces& \cite{MsbG,NIS2_2022,NIS2UmsuCG_2024,EU_AI_Act_2024}
       & \cite{VDE_LV_Rules,BSI_SMGW}  
       & \cite{EWI_BET_2025,BMWK_Schluesselmassnahmen_2025} \\
       c) Market- \& grid-oriented optimisation& \cite{EnWG,BNetzA.14a,BNetzA.14a_BK8,EEG,BNetzA_Redispatch_2020,ElectricityDirective2019,EU_AI_Act_2024}
       & \cite{VDE_FNN_Impuls_2024,VDE_LV_Rules,RD30,BMWK_Dataflex}
       & \cite{EWI_BET_2025,BMWK_Schluesselmassnahmen_2025} \\\hline
\end{tabular}
\label{tab:matrix}\\
\\

From this synthesis, several gaps and challenges emerge that motivate the subsequent system and roadmap discussions. 
Despite the variety of instruments, there is no fully integrated, LV-specific framework that jointly optimises measurement concepts, communication architectures and market- and grid-oriented optimisation methods under the given constraints, and data access rights, responsibilities and incentives for different actors (DSOs, metering point operators, suppliers, aggregators and customers) remain only partially specified for LV-relevant use cases. 
To pave the way towards active LV grid operation, Sec.~\ref{sec:system} introduces a structured system overview, which then serves as the basis for the research roadmap in Sec.~\ref{sec:roadmap}.

\vspace{-0.2em}
\section{Target system overview} \label{sec:system}
Enabling full active grid operation in LV grids is a core task for DSOs, but requires coordination between various roles and actors. 
Fig.~\ref{fig:system_overview} provides an overview of the main roles, communication paths and interfaces in the proposed target system. The cloud shapes refer to the pillars introduced in Sec.~\ref{sec:introduction} and indicate broad clusters of research and development needs.
\\
\begin{center}
    \includegraphics[width=0.48\textwidth]{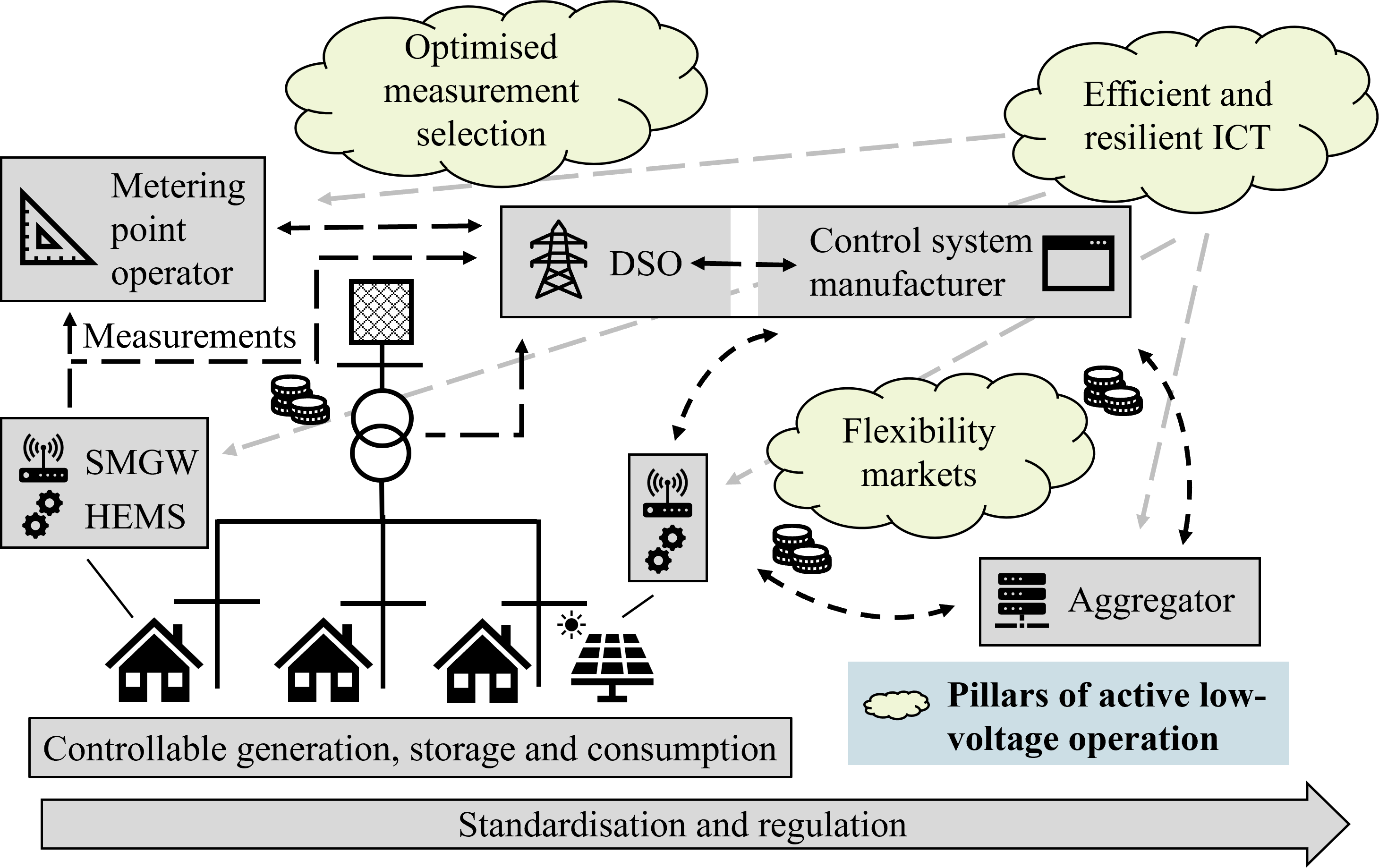} \label{fig:system_overview}
    
        \captionof{figure}{Components and their interaction in a fully active LV power system (own figure)}
        \label{fig:system_overview}
    \end{center}

As portrayed by the figure, controllable loads, generators and prosumers in LV grids constitute the basis for active operation. 
They are represented in Fig.~\ref{fig:system_overview} by households with home energy management systems and smart meter gateways and may bundle various controllable assets such as photovoltaic systems, electric vehicle chargers, heat pumps or battery storage systems. 
For DSOs, sufficient observability of their LV grids is necessary to determine preventive and curative measures. 
This requires reliable LV state estimation based on available measurement data. 
Measurements can stem from assets operated directly by the DSO, such as transformers and distribution feeder measurements, and from IMS. 
Depending on the specific data granularity, measurement data are often aggregated by metering point operators. 
With increasing smart meter roll-out, this raises the question for DSOs of how to select a cost-efficient set of measurements while still ensuring adequate state estimation quality. 
These functionalities are implemented in DSO control systems, typically supplied by specialised manufacturers. 
To reach the target system, in which measurement data are efficiently managed and processed, measurement selection algorithms and state estimation approaches suitable for modern and future LV grids are required.

The integration of market-based and grid-oriented optimisation is located at the intersection between grid users and DSOs. 
Aggregators can assume a coordinating role between private grid users and DSOs. 
While flexibility aggregation is often discussed for congestion management in higher voltage levels, market-based approaches for LV grids must balance aggregation with sufficient granularity to resolve local, grid-specific congestion in real time. 
In addition, hierarchical conflicts between market-based flexibility provision for higher voltage levels and flexibility use in LV grids must be avoided to reach the targeted active LV operation. 
The opportunities provided by current regulation should therefore be used to overcome local and temporal limitations of variable grid fees and dynamic tariffs.

Information and communication processes and infrastructure are essential to enable both measurement processing and flexibility activation. 
Resilient and efficient communication and interface architectures are a key requirement, especially as the amount of data to be aggregated and used by DSOs increases. 
Different data and system architectures need to be evaluated with respect to efficiency, security and privacy. 
This applies not only to the measurement side but also to flexibility aggregation and management, where efficiency, cybersecurity and resilience are essential to support the target system.

To realise active LV grid operation, DSOs must be able to develop and validate new solutions in dedicated research environments before deploying them in operational control systems.
This requires that algorithms can first be implemented and tested without time-consuming changes to proprietary control platforms, enabled by secure interfaces between DSO control-room software and external development and testing environments.
Such interfaces should support laboratory experiments, demonstration setups and field trials through standardised, flexible connections that meet stringent security requirements.
%To provide research contributions that can be implemented in practice at DSOs, it is necessary to initially develop and test solutions and algorithms without time-consuming changes and implementations in proprietary control systems. 
%Therefore, secure interfaces between DSO control room software and external development and testing environments are required. 
%These interfaces should enable laboratory settings, demonstration setups and field tests through standardised and flexible connections, while adequately addressing security requirements. 
While the research fields associated with the three main pillars mainly refer to technical developments and implementations, implications for regulation and standardisation must also be analysed across all research activities. 
Insights can inform adjustments and extensions to the regulatory framework discussed in Sec.~\ref{sec:regulation}.

\vspace{-0.2em}
\section{Roadmap} \label{sec:roadmap}
This section outlines the path to the target system in Sec.~\ref{sec:system}.
Existing research trends are first summarised (Sec.~\ref{sec:existing_activities}), then a roadmap is proposed, integrating all three pillars into operational practice (Sec.~\ref{sec:roadmap_}).

\subsection{Existing activities} \label{sec:existing_activities}
In recent research projects, important progress has been made towards active operation of distribution grids.
Building on the actor and interface model in Sec.~\ref{sec:system}, \cite{11180633} presents a prototyping framework for prosumer control strategies in LV grids that combines conceptualisation with simulations for grid planning and operation and models interactions among DSO, substation, HEMS and prosumers over SMGW-based communication paths. 
Using a grid-oriented price curve for EV charging as an application example, the framework demonstrates reductions in the occurrence and duration of operating limit violations and lower reinforcement needs in high penetration scenarios, while clarifying requirements for measurement granularity, data access and secure interfaces in line with the system overview in Sec.~\ref{sec:system}.
The authors of \cite{Zeng.2018} propose a low-carbon-based real-time electricity pricing mechanism that reflects the external characteristics of renewable energy and exploits demand-side flexibility. 
This pricing scheme is used as a control signal in a home energy management strategy to schedule flexible residential resources and illustrates how dynamic, signal-based control can contribute to market-based optimisation in LV grids.

% Distribution grid state estimation has been studied in a variety of works. 
On the state estimation side, the main challenge is how to derive the grid operating conditions from a limited number of measurements. 
AI-based approaches can be employed to estimate voltages or currents at unmeasured cables and buses, but their precision depends on available training data and grid models \cite{Jurczyk.2024}. 
Mathematical approaches such as Weighted Least Squares (WLS) are common in control room software, but they usually depend on the quality of pseudo-measurements used to complement real measurements. 
Novel approaches like in \cite{Pau_UnobservableWLS} offer alternative solutions to avoid the use of pseudo-measurements, but still lack full validation on real grids, for which a roadmap to field testing, validation, and possible roll-out is necessary. 
% Asman et al. \cite{Asman.2025} investigate how many measurements are needed to achieve a desired WLS estimation specificity and sensitivity. 
% They conclude that combining substation measurements with real-time data from IMS is beneficial for improving WLS estimates.

\subsection{Roadmap outline} \label{sec:roadmap_}

Building on the regulatory analysis and system overview, this section proposes a roadmap towards active operation of LV distribution grids. 
The roadmap is structured into four phases, as indicated by the rows, and each phase applies across the three main pillars introduced earlier, as shown in the columns of Fig. \ref{fig:roadmap}.
It is intended to bridge the gap between conceptual work and deployment in real distribution grids and to integrate existing research contributions outlined in Sec.~\ref{sec:existing_activities} into comprehensive system operation. \\

\begin{center}
    \vspace{-5mm}
    \includegraphics[width=0.48\textwidth]{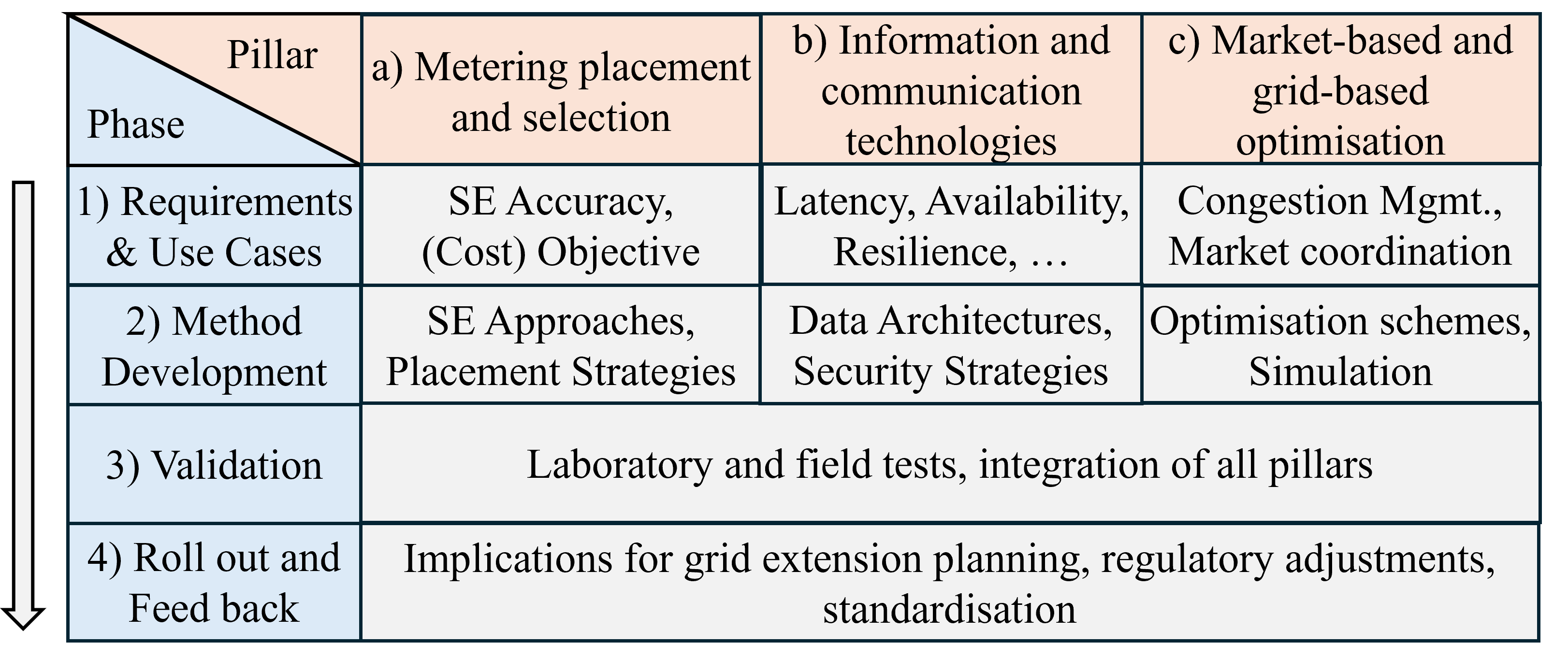} \label{fig:roadmap}
    
        \captionof{figure}{Roadmap towards active LV grids}
        \label{fig:roadmap}
    \end{center}

\textit{Phase 1)} focuses on requirements and use case definition. 
For measurement and observability, DSOs derive requirements for measurement types, spatial density, temporal resolution and data access from operational objectives such as state estimation accuracy, congestion management and operational optimisation. 
For communications and data management, requirements concern expected data volumes, latency bounds, availability and resilience targets, as well as role and rights concepts in line with the legal and cybersecurity framework. 
For market-based and grid-oriented optimisation, relevant use cases include local congestion management, mobilisation of flexibility from distributed generation, controllable demand and storage and coordination with existing market processes. 
Non-discrimination, transparency, security and computational tractability should be established as cross-cutting criteria in this phase.

\textit{Phase 2)} addresses method development and simulation. 
Within the first pillar, procedures are developed for cost-efficient placement of measurement points and for the selection of smart metering data under different optimisation objectives and grid characteristics, and classical state estimation methods and data-driven approaches are compared with respect to accuracy, robustness and data efficiency for representative urban and rural grids. 
In the second pillar, alternative data and system architectures are designed and assessed, including central and decentralised concepts and suitable communication protocols and security mechanisms, and tested in simulated scenarios that reflect realistic load and generation patterns and typical failure modes.
In the third pillar, market-based and grid-oriented optimisation schemes are formulated and analysed, including the design of local flexibility mechanisms and their hierarchical coordination with curative grid interventions as a fall-back and simulation studies on benchmark grids to assess congestion relief potential and cost efficiency.

\textit{Phase 3)} moves to laboratory and field validation. For the first pillar, methods for measurement selection and state estimation are tested with real or highly realistic measurement data, taking into account aggregation, missing values and delayed transmission. 
For the second pillar, communication and data infrastructures are implemented in laboratory environments and, where feasible, connected to existing control and metering systems, and stress tests with high data volumes, component outages and cyber incident scenarios support a systematic analysis of resilience and risk. 
For the third pillar, market-based and grid-oriented optimisation procedures are integrated with actual control room systems, operated in a non-intrusive mode without issuing real control commands to customer installations, to provide evidence on interoperability, computing times and the quality of operational decisions in realistic settings.

\textit{Phase 4)} addresses roll-out and system-level feedback. 
As active operation concepts mature, practical guidelines for measurement concepts, state estimation and operational processes in LV grids can be derived, and data from active operation are used to inform grid planning by quantifying the extent to which flexibility can defer or reduce conventional reinforcements and by identifying residual needs for grid expansion. 
For communications and data management, this phase leads to technical and organisational recommendations for secure and efficient information and communication technology architectures that comply with legal and standardisation requirements. 
For market-based and grid-oriented optimisation, recommendations are formulated on the design of local flexibility mechanisms, their integration with existing markets and congestion management regimes and the treatment of potential conflicts between local and system-wide uses of flexibility. 
Insights from all three pillars feed back into the refinement of requirements, regulatory frameworks and technical standards, thereby closing the iterative loop of the roadmap.

\vspace{-0.2em}
\section{Conclusion} \label{sec:conclusion}
The paper outlines a pathway towards active operation of low-voltage distribution grids. 
Starting from the German and European regulatory environment, it identifies three key pillars: measurement placement and observability, secure and efficient communications and interfaces and the integration of market-based and grid-oriented optimisation. 
It derives a structured system overview with key roles, interconnections and research topics and proposes a phased roadmap from requirements definition to roll-out. 
Future work focuses on implementing and evaluating these concepts in laboratory environments and real grids and on feeding the resulting insights back into grid planning, regulation and standardisation.

\vspace{-0.2em}
\section{Acknowledgements}
%FhGenie \cite{weber2024fhgeniecustomconfidentialitypreservingchat}, a customized Fraunhofer AI chatbot, was used to improve grammar and spelling in this paper. After using this tool, the authors reviewed and edited the content as needed. These tools were not used for text generation.
This work took inputs from the research projects "Redispatch 3.0" (FKZ 03EI4043I/J), "OptiQU" (FKZ 03EI4099A/C), "NI-PV 2030" (FKZ 03EE1231C)  and "OPium" (FKZ 03EIM4116), all funded by the Federal Ministry of Economic Affairs and Energy based on a decision by the German Bundestag.
The Fraunhofer AI chatbot FhGenie \cite{weber2024fhgeniecustomconfidentialitypreservingchat} was used for text improvement in this paper. The authors reviewed and edited all suggestions as needed and take full content responsibility.

\vspace{-0.2em}
\bibliographystyle{IEEEtran}
\bibliography{literature}

@misc{EEG,
  author = {{German Federal Ministry of Economic Affairs and Energy (BMWE)}},
  title  = {{Renewable Energy Sources Act ({EEG})}},
  year   = {2023}
}

@misc{EnWG,
  author = {{German Federal Ministry of Economic Affairs and Energy (BMWE)}},
  title  = {{Energy Industry Act ({EnWG})}},
  year   = {2005}
}

@misc{MsbG,
  author = {{German Federal Ministry of Economic Affairs and Energy (BMWE)}},
  title  = {{Act on the Operation of Metering Points ({MsbG})}},
  year   = {2016}
}

@techreport{BNetzA.14a,
  author      = {{Federal Network Agency (BNetzA)}},
  title       = {{Decision BK6-22-300: Adjustment of the Revenue Cap for Electricity Distribution System Operators}},
  institution = {Bundesnetzagentur (BNetzA)},
  year        = {2023},
  number      = {BK6-22-300},
  address     = {Bonn, Germany}
}

@techreport{BNetzA.14a_BK8,
  author      = {{Federal Network Agency (BNetzA)}},
  title       = {{Decision BK8-22-010: Network Charges for Controllable Connections and Consumer Equipment}},
  institution = {Bundesnetzagentur (BNetzA)},
  year        = {2023},
  number      = {BK8-22-010},
  address     = {Bonn, Germany}
}

@misc{NIS2_2022,
  author = {{European Parliament and Council of the European Union}},
  title  = {{Directive ({EU}) 2022/2555 on Measures for a High Common Level of Cybersecurity Across the Union ({NIS2})}},
  year   = {2022}
}

@misc{ElectricityDirective2019,
  author = {{European Parliament and Council of the European Union}},
  title  = {{Directive ({EU}) 2019/944 on Common Rules for the Internal Market for Electricity}},
  year   = {2019}
}

@misc{NIS2UmsuCG_2024,
  author = {{German Federal Ministry of the Interior and Community (BMI)}},
  title  = {{Act Reforming the {BSI} Act and Implementing the {NIS2} Directive ({NIS2UmsuCG})}},
  year   = {2024}
}

@techreport{BSI_SMGW,
  author      = {{Federal Office for Information Security (BSI)}},
  title       = {{Technical Guideline {BSI TR-03109}: Technical Communication Infrastructure for Intelligent Metering Systems}},
  institution = {Federal Office for Information Security (BSI)},
  year        = {2024}
}

@techreport{EWI_BET_2025,
  author      = {{EWI} and {BET}},
  title       = {{Energiewende. Effizient. Machen. Monitoring Report for the 21st Legislative Period}},
  institution = {EWI and BET},
  year        = {2025},
  address     = {Cologne/Aachen, Germany}
}

@misc{BMWK_Schluesselmassnahmen_2025,
  author      = {{Federal Ministry for Economic Affairs and Energy (BMWE)}},
  title       = {{Ten Key Measures Regarding the Monitoring Report: Becoming Climate Neutral while Remaining Competitive}},
  institution = {BMWK},
  year        = {2025},
  address     = {Berlin, Germany}
}

@techreport{BSI_Schutzprof,
  author      = {{Federal Office for Information Security (BSI)}},
  title       = {{Protection Profile for the Gateway of a Smart Metering System ({Smart-Meter-Gateway PP})}},
  institution = {Federal Office for Information Security (BSI)},
  year        = {2024}
}

@techreport{VDE_FNN_Impuls_2024,
  author      = {{VDE FNN}},
  title       = {{Impulse for Controlling Generation Systems in Low-Voltage Networks}},
  institution = {VDE FNN},
  year        = {2024},
  address     = {Berlin, Germany}
}

@techreport{BNetzA_Redispatch_2020,
  author      = {{Federal Network Agency (BNetzA)}},
  title       = {{Decision BK6-20-059 on the Determination of Conditions for Redispatch}},
  institution = {Bundesnetzagentur (BNetzA)},
  year        = {2020},
  number      = {BK6-20-059},
  address     = {Bonn, Germany}
}

@manual{RD30,
  author       = {{VDE}},
  title        = {{VDE SPEC 90032}: Redispatch 3.0 – Architecture and Processes for Congestion Management in Distribution Systems with Assets below 100 kW},
  organization = {VDE Association for Electrical, Electronic and Information Technologies},
  year         = {2025},
  address      = {Frankfurt am Main, Germany}
}

@inproceedings{Jurczyk.2024,
  author    = {Jurczyk, Kristina and Riedl, Leonie and Dipp, Marcel and others},
  title     = {{Comparing the Impact of AI-Based versus Standard Load Profiles in ANN State Estimation Training in a Real Distribution Grid}},
  booktitle = {Proc. Int. Conf. Smart Energy Systems and Technologies (SEST)},
  year      = {2024},
  pages     = {1--6},
  publisher = {IEEE}
}

@article{Zeng.2018,
  author  = {Zeng, Wei and von Appen, Jan and Selzam, Patrick and others},
  title   = {{Active Residential Load Management based on Dynamic Real-Time Electricity Price of Carbon Emission}},
  journal = {Energy Procedia},
  volume  = {152},
  pages   = {1027--1032},
  year    = {2018}
}

@inproceedings{11180633,
  author    = {Schoen, Andrea and Ringelstein, Jan and Mende, Denis and others},
  title     = {{Prototyping Control Strategies for Prosumers: A Framework for Strategy Development and Analysis in Low-Voltage Grids}},
  booktitle = {Proc. IEEE PowerTech},
  year      = {2025},
  pages     = {1--6},
  publisher = {IEEE}
}

@techreport{Heid2025_Future14aEnWG,
  author      = {Heid, Johannes and T{\"o}nges, Eric and Wende-von Berg, Sebastian and others},
  title       = {{The Future of Section 14a EnWG: A Roadmap towards Active Operation of Low-Voltage Grids}},
  type        = {White paper},
  institution = {Fraunhofer CINES},
  year        = {2025}
}

@online{BMWK_Dataflex,
  author = {{Federal Ministry for Economic Affairs and Energy (BMWE)}},
  title  = {{Manufacturing-X: Project Dataflex}},
  year   = {2026},
  url    = {https://www.bundeswirtschaftsministerium.de/Redaktion/DE/Dossier/Manufacturing-x/Module/projekt-dataflex.html},
  note   = {in German}
}

@misc{EU_AI_Act_2024,
  author = {{European Union}},
  title  = {{Regulation ({EU}) 2024/1689 on Harmonised Rules on Artificial Intelligence ({AI Act})}},
  year   = {2024}
}

@misc{weber2024fhgeniecustomconfidentialitypreservingchat,
      title={{FhGenie: A Custom, Confidentiality-preserving Chat AI for Corporate and Scientific Use}}, 
      author={Ingo Weber and Hendrik Linka and Daniel Mertens and others},
      year={2024},
      eprint={2403.00039},
      archivePrefix={arXiv},
      primaryClass={cs.SE},
}

@techreport{VDE_LV_Rules,
  author      = {{VDE FNN}},
  title       = {Technical Connection Rules for the Low-Voltage Network ({VDE-AR-N 4100} and {VDE-AR-N 4105})},
  institution = {Verband der Elektrotechnik Elektronik Informationstechnik e.V. (VDE)},
  address     = {Berlin, Germany},
  year        = {2018/2019},
  note        = {Comprehensive framework for connection and operation of loads and generators in German low-voltage grids},
  language    = {German}
}

@ARTICLE{Pau_UnobservableWLS,
  author={Pau, Marco and Pegoraro, Paolo Attilio},
  journal={IEEE Transactions on Instrumentation and Measurement}, 
  title={WLS-Based State Estimation for Unobservable Distribution Grids Through Allocation Factors Evaluation}, 
  year={2024},
  volume={73},
  number={},
  pages={1-13},
  doi={10.1109/TIM.2024.3387498}}

\end{multicols}
\end{document}